\documentclass[twocolumn,prd,secnumarabic,
amsmath,amssymb,superscriptaddress,nofootinbib]
{revtex4-2}

\usepackage{lineno,hyperref}
\usepackage{graphicx}
\usepackage{bm}
\usepackage{amsmath} 
\usepackage{amssymb}
\usepackage{xcolor}

\newcommand{\mL}{{\mathcal{L}}}
\newcommand{\mE}{{\mathcal{E}}}
\newcommand{\mB}{{\mathcal{B}}}
\newcommand{\mD}{{\mathcal{D}}}
\newcommand{\mH}{{\mathcal{H}}}
\newcommand{\mP}{{\mathcal{P}}}

\newcommand{\req}[1]{Eq.\,(\ref{#1})} 
\newcommand{\rs}[1]{section~\ref{#1}}

\begin{document}

\title{Everlasting interaction: polarization summation without a Landau pole} 

\author{Stefan Evans}
\affiliation{Department of Physics, The University of Arizona, Tucson, AZ 85721, USA}
\affiliation{Helmholtz-Zentrum Dresden-Rossendorf, Bautzner Landstraße 400, 01328 Dresden, Germany}
\author{Johann Rafelski}
\affiliation{Department of Physics, The University of Arizona, Tucson, AZ 85721, USA}

\begin{abstract}
We propose an external field approach to evaluating effective action allowing the interaction to act everywhere at all times (everlasting).  Requiring that the asymptotic gauge fields are always-interacting, we implement displacement fields encoding polarization corrections into the derivation of effective action. The result is a novel polarization summation for one-cut reducible loop diagrams, which can be applied to two cases: transient quasi-constant electromagnetic fields, and everlasting interactions. In the first case, a perturbative expansion of our result recovers the Schwinger-Dyson reducible diagram series with a Landau pole. The everlasting summation evaluated in nonperturbative fashion removes the Landau pole, providing a new avenue for modeling strongly interacting theories. 
\end{abstract}

\maketitle

\section{Introduction}

Among the unresolved mysteries in quantum field theory is the Landau pole~\cite{Landau:1955nc, Huang:2013zaa} -- an unexpected singular point in the sum arising in the vacuum polarization chain-diagrams.  The perturbative QED framework suggests this result arises considering  the interaction between photons and electron-positron loop fluctuations in a manner akin to a scattering problem with switch-on-off procedure, with the Schwinger-Dyson equations~\cite{Dyson:1949ha, Schwinger:1951ex, Schwinger:1951hq} built upon bare asymptotic photon states. Experimental tests of this suggested pole remain elusive however, given the small QED coupling. In QCD on the other hand, there is strong evidence for the absence of the Landau pole~\cite{Fischer:2006ub, Aguilar:2008xm}, supported  by lattice computation and  measurements of the running coupling~\cite{Binosi:2016nme, Deur:2016tte, Deur:2022msf, Ding:2022ows, Deur:2023dzc, Ferreira:2023fva}.

Here we explore the question of the Landau pole 
using an external field method for computing effective action in quasi-constant (infrared) fields. A convenient theoretical environment to develop our model example is the electromagnetic (EM) vacuum response considered in the context of the Euler-Heisenberg-Schwinger (EHS) effective action~\cite{Heisenberg:1935qt,Weisskopf:1996bu,Schwinger:1951nm,Dunne:2004nc}. In this context the nature of the interaction of the fields was remarked on by Weisskopf~\cite{Weisskopf:1996bu}:  \emph {\lq\lq One can by no means separate the external field from the field that is created by the vacuum electrons themselves\rq\rq}.

Taking inspiration from  Weisskopf's remark, we distinguish the perturbative switch on/off approach to evaluating effective action, in which the external fields are asymptotically non-interacting far from the interaction region, from what we propose to call hereon as \lq everlasting\rq\  interactions. In the latter case, the gauge field cannot be removed from the interaction region. We develop the everlasting approach by implementing displacement (distinct from EM) fields to encode the polarization effects in a self-consistent manner. The result for our illustrative example built on the EHS action is a one-cut reducible loop summation which exhibits, near the expected Landau pole where the interaction strength is large, a difference in outcomes compared to the usual perturbative approach.

Our extension of the EHS effective action may be seen as a demonstration of the everlasting principle, to be implemented in consideration of any other QFT effective action in which the asymptotic fields are always interacting. We focus on the polarization function and the Landau pole as it provides a model for studying more strongly interacting theories, including as an interesting example the Savvidy Yang Mills vacuum~\cite{Savvidy:1977as, Matinyan:1976mp, Nielsen:1978rm, Savvidy:2019grj, Cho:2000ck, Ozaki:2015yja, Savvidy:2022jcr, Savvidy:2023aa}.  The strongly coupled high mass sector of electroweak theory e.g. the strong coupling of the Higgs boson to the top quark, and W, Z bosons~\cite{Labun:2012fg, ATLAS:2012yve, CMS:2012qbp, Bernreuther:2008ju, Shifman:1979eb, Marciano:2011gm} offers another domain of future application of the everlasting principle in study of effective action.

\section{Motivation}
\label{secMotivation}

Weisskopf's insight, that the externally applied field entering the EHS action has to be made consistent with the polarized vacuum, implies the need to sum higher order one-cut reducible loop diagrams. Clearly any effect of such a higher order summation would be felt in presence of strongly coupled or very strong fields. 

In perturbative formulation for quasi-constant fields  of infinite extent the  related diagrams starting with the two-loop reducible action were, following Ritus~\cite{Ritus:1975cf}, for a long time assumed to vanish. Recently however, Gies and Karbstein~\cite{Gies:2016yaa} discovered that in the limit of vanishing momentum (quasi-constant fields), the vanishing factor in the Ritus argument is compensated by the virtual photon propagator connecting the two loops, producing a nonzero result. 

As a step in the direction of Weisskopf's observation, 
the field-theoretical approach in~\cite{Gies:2016yaa} inserts a dynamical EM field correction to the external background. This effectively encodes perturbative interactions with the polarized vacuum, with electron loop degrees of freedom integrated out.

The effective EM action, including the proposed two-loop complement to the EHS result~\cite{Gies:2016yaa} reads as
\begin{align}
\mL_{\mathrm{eff}}^{\rm GK}=&\;\mL_\mathrm{M}+\mL_1+\mL_2
\;,
\end{align}
with the Maxwell term
\begin{align}
\mL_\mathrm{M}=&\;\frac{\mE^2-\mB^2}2=-\frac{F^{\mu\nu}F_{\mu\nu}}4
\end{align}
in terms of EM tensor $F^{\mu\nu}$. The one-loop EHS action
\begin{align}
\label{pert1}
\mL_1=\mL_\mathrm{M}\Pi_0+ \mL_1^\mathrm{r}
\;,
\end{align}
where $\Pi_0$ is the logarithmically divergent vacuum polarization to be removed by charge renormalization, and $\mL_1^\mathrm{r}$ is the renormalized nonlinear (fourth order and higher in EM field) EHS contribution, discussed in more detail below.
The two-loop contribution Eq.\,(32) of~\cite{Gies:2016yaa} reads 
\begin{align}
\label{pert2}
\mL_2=\frac12\frac{\partial \mL_1^\mathrm{r}}{\partial F^{\mu\nu}}\frac{\partial \mL_1^\mathrm{r}}{\partial F_{\mu\nu}}
\;,
\end{align}
originally evaluated using the renormalized (finite) contribution to the one-loop action~\footnote{
A factor $-1/2$ difference with respect to the generally accepted Schwinger-Dyson series at two-loop order arises, which propagates into  sequel work higher order loop extensions~\cite{Karbstein:2019wmj}, see~\cite{Evans:2024ddy}.}

In this work we develop an alternate external field method, with polarization effects encoded as part of the asymptotic fields in a self-consistent manner. Aside from sign and magnitude considerations, see~\cite{Evans:2024ddy}, we obtain a loop summation action in concordance with the EM derivative structure of~\req{pert2} and sequel perturbative higher order loop extensions~\cite{Karbstein:2019wmj}. Implementation to all orders of our procedure  removes the Landau pole which is the key result we report here.

\section{Everlasting field interaction}

In the EHS action framework, only the electron field is 2nd-quantized in our approach and thus we sum diagrams which do not involve \lq internal\rq\ photon lines (inside an electron loop). However, there are infinite (reducible) vacuum polarization diagrams that can be derived as a polarization effect, without need for photon field quantization and integration over virtual photon momentum.

To develop our approach we first show the non-interacting case as it yields the EHS one-loop result. We then implement always-interacting external fields, producing a differential equation with the one-loop expression as an input function.

\subsection{Non-interacting case}

Consider a charged spin-1/2 Dirac particle, thus a source of the electrical field $\mE_{\mathrm{e}}$ (e for electron), entering an external constant electrical field $\mE_{\mathrm{X}}$. To describe their interaction we write the EM Maxwell action for both fields, with the remainder $ \mL_{\mathrm{Dirac}}=\bar\psi(\gamma\cdot p- m)\psi$:
\begin{align}
\label{interaction}
W=&\,\int d^4x (\mL_\mathrm{M}+\mL_{\mathrm{Dirac}})
\nonumber \\
=&\;
\frac12\int d^4x (\mE_{\mathrm{X}}+\mE_{\mathrm{e}})^2
+\int d^4x\bar\psi(\gamma\cdot p- m)\psi
\nonumber \\
=&\;\frac12\int d^4x (\mE_{\mathrm{X}}^2+\mE_{\mathrm{e}}^2)
\nonumber \\
&\;
+\int d^4x[ \mE_{\mathrm{X}}\cdot\mE_{\mathrm{e}}+\bar\psi(\gamma\cdot p- m)\psi]
\;.
\end{align} 
In the last line we combined the mixed-field term, producing the interaction between the particle and field, with the particle action. Upon integration by parts
\begin{align}
\label{intterm}
\int d^4x \mE_{\mathrm{X}}\mE_{\mathrm{e}}
=&\,-\int d^4x (\nabla A^0_{\mathrm{X}})\mE_{\mathrm{e}}
\nonumber \\
=&\,
\int d^4x A^0_{\mathrm{X}}\rho_{\mathrm{e}}
=\int d^4x \bar\psi A^0_{\mathrm{X}}\gamma_0\psi
\;,
\end{align}
where $\rho_{\mathrm{e}}=\psi^\dagger\psi$ and $\bar\psi=\psi^\dagger\gamma_0$. 
The surface terms vanish due to charge conservation as imposed by gauge invariance: this is also seen considering $\mE_{\mathrm{e}}$ and its derivatives describe a single localized particle fluctuation. The remaining two field terms in Eq.\,(\ref{interaction}) describe the field action of the classical field and the classical electron self-energy.

Inserting Eq.\,(\ref{intterm}) into last line of Eq.\,(\ref{interaction}), we obtain the action for the Dirac field in the presence of an external $A^0_{\mathrm{X}}$-potential. Applying covariance argument we generalize the potential to a full four-vector $eA^\mu_{\mathrm{X}}$, replacing $p^\mu \to p^\mu-eA^\mu_{\mathrm{X}}$. 
Upon 2nd-quantization of the Dirac field one computes the EHS action function for constant fields generated by potential $A^\mu_{\mathrm{X}}$~\cite{Weisskopf:1996bu,Heisenberg:1935qt,Schwinger:1951nm,Dunne:2004nc}:
\begin{align}
\label{L1form0}
\mL_1=\!\!\!&\;\int\! d^4x \langle 0|\bar\psi(\gamma\cdot( p-eA_{\mathrm{X}})- m)\psi +\mathrm{h.c.}|0\rangle 
\;,
\end{align}
to obtain~\req{pert1}:
\begin{align}
\label{L1form01}
\mL_1 =&\;
\frac{\mE_{\mathrm{X}}^2}2\Pi_0+ \mL_1^\mathrm{r}\;,
\qquad
\mL_1^\mathrm{r}=\mathcal{O}(\mE_{\mathrm{X}}^4)
\;.
\end{align}

Having summarized above the one-loop EHS action based on non-interacting fields, we now develop an illustrative example of asymptotically interacting fields. This amounts to the effective action~\req{L1form01}, describing the vacuum interaction, feeding back into (polarizing) the external field $\mE_{\mathrm{X}}$ prescribed at the start in~\req{interaction}.

\subsection{Always-interacting case}

As seen above, the vacuum acquires, through the evaluation of the field-dependent action function~\req{L1form0}, the properties of a dielectric. This dielectric behavior is an everlasting in time and infinitely-spanning in all space vacuum state in which the fields exist. We can no longer assume an a priori prescribed Maxwell electromagnetic Lagrangian $\mL_\mathrm{M}$ as in~\req{interaction}, based on non-interacting fields. To account for fields being already polarized by the  vacuum we introduce $\mL_{\mathrm{eff}}$ as the self-consistent effective EM Lagrangian: 
\begin{align}
\label{separate}
\Big(\mL_\mathrm{M}=\mE^2/2\Big)
\to \mL_{\mathrm{eff}}
\;.
\end{align}
This effective action describes reducible loop polarization effects where the electron degrees of freedom are a priori integrated out.

Our objective is to develop the interaction term between the polarized background field and a single electron fluctuation. Since the vacuum dielectric response must be solved for self-consistently, we expand in powers of the electron sourced field $\mE_{\mathrm{e}}$, of negligible magnitude compared to a prescribed external field $\mE_{\mathrm{X}}$. We expand in the small fluctuation to obtain 
\begin{align}
\label{interaction2}
\tilde W=&\,\int d^4x (\mL_{\mathrm{eff}}(\mE)
+  \mL_{\mathrm{Dirac}})
 \\ \nonumber
=&\,\int d^4x \Big(\mL_{\mathrm{eff}}(\mE)|_{\mE_{\mathrm{X}}}
+\frac12\!\!\left.\frac{\partial^2\mL_{\mathrm{eff}}(\mE )}
  {\partial\mE_{\mathrm{i}}\partial\mE_{\mathrm{j}}}\right\vert_{\mE_{\mathrm{X}}}\!\!\!
   \mE_{\mathrm{e}}^{\mathrm{i}}\mE_{\mathrm{e}}^{\mathrm{j}}
+...\Big)
 \\ \nonumber
\hphantom=&\,
+\int d^4x\Big(\left.\frac{\partial\mL_{\mathrm{eff}}(\mE)}
    {\partial\mE_{\mathrm{i}}}\right\vert_{\mE_{\mathrm{X}}}\!\!\! \mE_{\mathrm{e}}^{\mathrm{i}}
+\bar\psi(\gamma\cdot p- m)\psi\Big)
\;.
\end{align}

In the leading term of the expansion we recognize the displacement field $\mD_{\mathrm{X}}$:
\begin{align}
&\,\mE_{\mathrm{X}} \to \mD_{\mathrm{X}}=\frac{\partial\mL_{\mathrm{eff}}}{\partial\mE_{\mathrm{X}}}
\;,
\end{align}
where polarization contributions $\mP$ enter the displacement field according to~\cite{BialynickiBirula:1984tx}
\begin{align}
\label{Pform}
\mD_{\mathrm{X}}=&\;\mE_{\mathrm{X}}+\mP
\;.
\end{align}
We identify the interaction term coupling to probe charge and integrate by parts: 
\begin{align}
\label{Atildeform}
\int d^4x \frac{\partial\mL_{\mathrm{eff}}(\mE_{\mathrm{X}})}{\partial\mE_{\mathrm{X}}}\mE_{\mathrm{e}}
=&\,-\int d^4x (\nabla \tilde A^0_{\mathrm{X}})\mE_{\mathrm{e}}
\nonumber \\
=&\,
\int d^4x \tilde A^0_{\mathrm{X}}\rho_{\mathrm{e}}
\;,
\end{align}
with surface terms canceling again due to charge conservation, also noting $\mE_{\mathrm{e}}$ describes a single localized fluctuation like before. However, rather than seeing a non-interacting field, the electron is subject to the external field which a priori encodes the everlasting interaction with the collective (infinite) fluctuations spanning the vacuum. This field is described by potential $\tilde A^\mu_{\mathrm{X}}$, defined as a basis for the displacement field. This potential can be recombined with  the Dirac particle action. 

The derivation of effective action proceeds like in the EHS approach resulting in the effective nonlinear action~\cite{Weisskopf:1996bu,Heisenberg:1935qt,Schwinger:1951nm,Dunne:2004nc}. The same one-loop functional dependence emerges -- except that now the gradient of this potential in~\req{Atildeform} enters, producing in the evaluation of the effective action the displacement field. This is the key difference -- our consideration allows for the existence of effective nonlinear action ab-initio, in comparison to the usual perturbative approach. We return to the effect this has on the EHS action and on the electron self-energy under separate cover. These problems are nonlinear and more intricate, while the study of the Landau pole has an analytical solution.

Therefore here we focus our attention on how these considerations impact the charge renormalization. For this we consider the quadratic in EM field term arising in effective action, the one-loop function. This term corrects the Maxwellian term   
\begin{align}
\label{orignalApproach}
\mL_{\mathrm{eff}}(\mE)=&\;
\frac{\mE^2}2+\mL_1\Big(\frac{\partial\mL_{\mathrm{eff}}(\mE)}{\partial\mE}\Big)
\;, 
\end{align}
where above and from hereon we have dropped the label $X$ in the subscript. According to \req{L1form01} to leading order in EM fields, keeping the quadratic in EM field contribution to charge renormalization the EHS action function is 
\begin{align}
\label{L1form}
 \mL_1(z)=\frac{z^2}2\Pi_0+\mathcal{O}(z^4) 
\;.
\end{align}
 
These~\req{orignalApproach} and~\req{L1form} create a nested differential equation, with the one-loop function $\mL_1$ as input: 
\begin{align}
\label{orignalApproach2}
\mL_{\mathrm{eff}}(\mE)=&\;
\frac{\mE^2}2+
\frac{\Pi_0}2\Big(\frac{\partial\mL_{\mathrm{eff}}(\mE)}{\partial\mE}\Big)^2+
\mathcal{O}\Big(\frac{\partial\mL_{\mathrm{eff}}(\mE)}{\partial\mE}\Big)^4
\;.
\end{align} 
Our approach encodes the always-interacting field structure within its argument ($\partial\mL_{\mathrm{eff}}/\partial\mE$). This allows us to unravel the nonperturbative structure by solving for the relation between $\mE$ and $\partial\mL_{\mathrm{eff}}/\partial\mE$.

\section{Continuous fraction vacuum response}
\subsection{Vacuum polarization and the Landau pole}
\label{vacPol41}
Since we consider in the study of the Landau pole the charge renormalizing contribution to vacuum response, the differential equation form of effective action in~\req{orignalApproach2} is analytically solvable. More generally however, keeping the nonlinear terms (light-light scattering and higher orders in~\req{L1form}) will likely require numerical solutions, which we will address under separate cover. 

We can write the solution with an expression that is quadratic in EM fields, times a constant $(1+\Pi_{\mathrm{eff}})$:
\begin{align}
\label{fullform}
\mL_{\mathrm{eff}}(\mE)
\equiv&\;
\frac{\mE^2}2\Big(1+\Pi_{\mathrm{eff}}\Big)
\;.
\end{align}
$\Pi_{\mathrm{eff}}$ describes the polarization response -- the object of interest which we set out to compute in order to describe the self-consistent dressed fields.

To solve for $\Pi_{\mathrm{eff}}$, we plug~\req{fullform} and its derivative with respect to $\mE$ into~\req{orignalApproach2}, so that the  polarization response (\req{Pform}) modifies the field entering into the effective action function $\mL_1$:
\begin{align}
\label{fullform2}
\frac{\mE^2}2\Big(1+\Pi_{\mathrm{eff}}\Big)
=&\;
\frac{\mE^2}2+\mL_1\Big(\frac{\partial\mL_{\mathrm{eff}}}{\partial \mE}\Big)
\\ \nonumber
=&\;
\frac{\mE^2}2+\mL_1\Big(\mE(1+\Pi_{\mathrm{eff}})\Big)
\;.
\end{align}
$\Pi_{\mathrm{eff}}$ appears on both sides of~\req{fullform2} in a nested expression. Applying the one-loop function $\mL_1$ from~\req{L1form}, the quadratic in EM field dependence cancels, and after some algebra we obtain 
\begin{align}
\label{inputPol}
1+\Pi_{\mathrm{eff}}=\frac1{1-\Pi_0(1+\Pi_{\mathrm{eff}})}
\;.
\end{align}
We recognize a Schwinger-Dyson-like summation, with the key distinction being that the polarization function $\Pi_0$ gains an additional factor $(1+\Pi_{\mathrm{eff}})$. This factor iterates as a continuous fraction
\begin{align}
\label{Nesteda}
1+\Pi_{\mathrm{eff}}=&\;
\frac{1}{
1-\dfrac{\Pi_0}{
 1-\dfrac{\Pi_0}{
  1-\dfrac{\Pi_0}{
   1-\cdots
   }
  }
 }
}
\;.
\end{align}

In this exceedingly simple case there is also an analytical solution which can be inferred directly from~\req{inputPol}:
\begin{align} 
\label{PsumFulla}
1+\Pi_\mathrm{eff}=&\;
\frac{1}{1/2+\sqrt{1/4-\Pi_0}}
\;.
\end{align}
Moreover, we recall that the function replacing the polarization function entering the Schwinger-Dyson equation is 
\begin{align} 
\label{PsumFullaB}
\Pi_0(1+\Pi_\mathrm{eff})=&\;
 {1/2-\sqrt{1/4-\Pi_0}}
\;.
\end{align}
The Landau pole requires that there is a zero in 
\begin{align} 
1-\Pi_0(1+\Pi_\mathrm{eff})=&\;
 1/2+{\sqrt{1/4-\Pi_0}}\ne 0
\;,
\end{align}
which clearly as indicated cannot ever happen for the physical form of $\Pi_0$,
\begin{align} 
\label{Pi0form}
\Pi_0=\frac{e^2}{12\pi^2}\Big(\delta^{-1}-\gamma_E-\ln(m^2)\Big)
\;,
\end{align}
with $\delta^{-1}$ following from dimensional regularization and relating to cutoff $\Lambda$-dependence as $\ln(\Lambda^2)$. This form suggests that instead of a Landau pole, we ultimately at short distances will encounter finite renormalization, or branch cuts, depending on the theory studied. These solutions do not necessarily have the character of a freely propagating particle - we hope that their understanding will arise in the future. 

One can also look at the perturbative in $\Pi_0$ expansion:
\begin{align} 
\label{Peff}
1+\Pi_\mathrm{eff}\underset{\Pi_0\ll 1}=
1+\Pi_0+2\Pi_0^2+5\Pi_0^3+14\Pi_0^4+\cdots 
\;,
\end{align}
where we note that departure from the perturbative Schwinger-Dyson series,
\begin{align} 
\label{PSD}
\frac{1}{1-\Pi_0}\underset{\Pi_0\ll 1}=
1+\Pi_0+\Pi_0^2+\Pi_0^3+\Pi_0^4+\cdots 
\;,
\end{align}
begins at the two-loop order coefficients.  However, our primary result is significant when the effective strength of interaction is large, thus  at very large $q^2$. We next present a specific example for the case of strong magnetic fields which attracted attention in other works.

\subsection{Strong magnetic fields}
\label{strongBexample}
We now obtain the effective action in the everlasting interaction case for a pure magnetic constant field. We repeat the steps in~\req{interaction2}, where the self-consistent polarization corrections to the magnetic field amount to using the displacement field 
\begin{align}
\label{Bdiffform}
\mH(\mB)=-\frac{\partial\mL_{\rm eff}}{\partial\mB}
\;
\end{align}
in the argument of the one loop EHS action. The resulting differential equation has now the form
\begin{align}
\label{orignalApproachB}
\mL_{\mathrm{eff}}(\mB)=&\;
-\frac{\mB^2}2+\mL_1\Big(-\frac{\partial\mL_{\mathrm{eff}}(\mB)}{\partial\mB}\Big)
\;.
\end{align}

To compare with other works, we take the strong $\mB$ limit of the renormalized one loop EHS action
\begin{align}
\label{EHSBlim}
\lim_{e\mB/m^2\gg1}\mL_1^{\rm r}(\mB)=&\;-\frac{\mB^2}2\Pi_\mB
\;,
\end{align}
where polarization $\Pi_\mB$ is given by 
\begin{align}
\label{PiBform}
\Pi_\mB\equiv&\; -\frac{e^2}{12\pi^2}\ln\Big(\frac{e\mB}{m^2}\Big)
\;.
\end{align}

A single chain of reducible loop diagrams dominates in the strong $\mB$ limit, as evidenced by the derivative 
\begin{align}
\lim_{e\mB/m^2\gg1}-\frac{\partial\mL_1^{\rm r}}{\partial\mB}=&\; \mB\Pi_\mB +\frac{\mB^2}2\frac{\partial\Pi_\mB}{\partial\mB}
\approx \mB\Pi_\mB
\;,
\end{align}
a feature noted by Karbstein~\cite{Karbstein:2019wmj}. We treat the magnetic field inside  $\Pi_\mB$ as a constant. In consequence given all approximations each electron loop couples to a maximum of two (reducible) virtual photons resulting in a single chain continued fraction sum. Therefore all steps we presented in the study of the Landau pole apply. 

Repeating the steps from~\rs{vacPol41} with a constant $\Pi_\mB$ we can write the effective action as a quadratic in $\mB$ solution, rendering the differential equation~\req{orignalApproachB} analytically solvable:
\begin{align}
\label{fullformB}
\mL_{\mathrm{eff}}(\mB)
=&\;
-\frac{\mB^2}2\Big(1+\Pi_{\mathrm{eff}}\Big)
\\ \nonumber
=&\;
-\frac{\mB^2}2+\mL_1^{\rm r}\Big(\mB(1+\Pi_{\mathrm{eff}})\Big)
\;.
\end{align}
Applying $\mL_1^{\rm r}$ from~\req{EHSBlim}, we obtain 
\begin{align}
\label{NestedstrongB}
1+\Pi_{\mathrm{eff}}
=&\;
\frac1{1-\Pi_\mB(1+\Pi_{\mathrm{eff}})}
=
\frac{1}{
1-\dfrac{\Pi_\mB}{
 1-\dfrac{\Pi_\mB}{
  1-\dfrac{\Pi_\mB}{
   1-\cdots
   }
  }
 }
}
\;.
\end{align}
Recalling that the series~\req{NestedstrongB} follows~\req{inputPol}, which can be expressed as a square root and a series expansion in powers of loop order $\Pi_\mB$,
\begin{align} 
\label{SQRTstrongB}
\mL_{\mathrm{eff}}(\mB)
 =&\;
-\frac{\mB^2}2
\frac{1-\sqrt{1 -4\Pi_\mB}}{2\Pi_\mB}
\\[0.1cm] \nonumber
 =&\;
 -\frac{\mB^2}2\Big(1+\Pi_\mB+2\Pi_\mB^2+5\Pi_\mB^3+{\mathcal{O}}(\Pi_\mB^4)\Big)
\;.
\end{align}

We recall our everlasting interaction approach to developing the above expressions is based on asymptotically interacting fields that exist at all times. On the other hand in context of strong EM field environments, it is important to note that all EM fields occurring in nature span finite spacetime domains. 

To apply our result to such switch on/off EM fields, we expand the RHS of~\req{NestedstrongB}, this time in powers of $\Pi_{\rm eff}$:
\begin{align}
\label{truncatedfom}
\frac1{1-\Pi_\mB(1+\Pi_{\mathrm{eff}})}
=\frac1{1-\Pi_\mB}
+\frac{\Pi_{\mathrm{eff}}}{(1-\Pi_\mB)^2}
+{\mathcal{O}}(\Pi_{\mathrm{eff}}^2)
\;.
\end{align}
Truncating~\req{truncatedfom} to the first term (0th order in $\Pi_{\mathrm{eff}}$), we recover the perturbative switch on/off polarization series (\req{PSD} with $\Pi_\mB$ in place of $\Pi_0$) built on asymptotically non-interacting external fields. This perturbative form can be compared to the prior reducible loop summations~\cite{Karbstein:2019wmj, Karbstein:2017gsb, Karbstein:2021gdi} extending the two-loop action in~\cite{Gies:2016yaa}, also supported by our earlier consideration~\cite{Evans:2023jqy}. Our result preserves the generally accepted sign of the Schwinger-Dyson series at two-loop and higher orders, resolving prior sign and magnitude prefactors~\cite{Evans:2024ddy}.

Interestingly the everlasting interaction result expressed as a continuous fraction in~\req{NestedstrongB}, i.e. ~\req{truncatedfom} to all orders in $\Pi_{\mathrm{eff}}$,  is also a single cut reducible loop sum in the limit of large magnetic fields. The key difference is that  there is no  Landau pole, irrespective of the sign of $\Pi_\mB$ as seen in~\req{SQRTstrongB}. This shows the potential for application of our continued fraction summation to strongly interacting chromo-magnetic fields and the Savvidy~\cite{Savvidy:1977as,  Savvidy:2022jcr, Savvidy:2023aa} vacuum state.

\section{Conclusions and outlook}

In the context of the external field method for evaluating effective action, we took inspiration from Weisskopf's comment to consistently define the meaning of the electric and magnetic fields in the absence of an asymptotic region where these fields do not experience polarization effects. Our procedure consists of: 1) Recognition of external fields entering the  one-loop effective action derivation to be the displacement fields; and 2) Implementation of nonperturbative summation before an attempt to define renormalized charge. Our result is distinct from the usual perturbative method based on asymptotically non-interacting fields. 

In this manner, we encoded nonperturbative reducible loop corrections into the standard  approach to deriving one-loop effective action, using EHS as an example. We demonstrated in~\rs{vacPol41} that the Schwinger-Dyson series transforms upon resummation into a continuous fraction expression for effective polarization $\Pi_{\mathrm{eff}}$,~\req{Nesteda}. This contains no Landau pole: at the value of one-loop polarization $\Pi_0$ at which one normally expects a pole, the summed renormalization constant remains finite. However, at large enough $\Pi_0$, the improved $\Pi_{\mathrm{eff}}$ can become complex -- at this point the driving $\Pi_0$ function needs further exploration.

In~\rs{strongBexample} we evaluated the effective action in strong magnetic field backgrounds. We applied our everlasting approach, based on infinitely spanning in time fields, to realistic fields that span finite spacetime domains using perturbative expansion of the action in~\req{truncatedfom}. From this expansion we recovered the structure of the perturbative reducible diagram summation proposed in prior works, while resolving sign and magnitude prefactors to ensure consistency with the Schwinger-Dyson series~\cite{Evans:2024ddy}. To our knowledge this is the first derivation of reducible loop diagrams in strong fields stemming from the exceedingly simple recognition that the asymptotically interacting fields in the effective action are the displacement fields.

Up to this point we have considered closed electron loops; reducible diagram summations were recently obtained for  EHS loop contributions to spin-0~\cite{Edwards:2017bte} and spin-1/2 propagators~\cite{Ahmadiniaz:2017rrk}. Moreover, extensions of the summation procedure to different field configurations beyond the constant field EHS limit are possible~\cite{Ahmadiniaz:2019nhk}.

We recall that in the external field approach in~\req{interaction2} we have truncated the expansion at lowest power of electron field. Higher powers in $\mE_{\mathrm{e}}$ e.g. corresponding to two and higher photon cut reducible diagrams require further consideration. Some diagram classes are relating to the self-energy of the probing particle; these contributions show similarities with mass catalysis~\cite{Gusynin:1995gt,Gusynin:1999ti,Ferrer:2000ed,Elizalde:2002ca,Shovkovy:2012zn}. It is important to note in this context that the mathematical approach presented here needs to be tested in many regimes and will perhaps evolve further.

To obtain the full polarization summation beyond one-cut reducible diagrams, one needs to precisely account for all terms in $\Pi_0$, see \req{Peff}. For example we need to incorporate an internal photon to obtain up to second order polarization effects. To be exact to 3rd order one would also need to incorporate higher order cut reducible diagrams. Such corrections in higher order incorporate their own continued fraction summation. We mention this in order to clarify that a systematic study of the everlasting nonperturbative vacuum structure reaches far beyond the usual polarization series. This clarifies why the current study is focused on the Landau pole.

In summary: we have proposed and developed a novel external field approach, implementing displacement fields in the derivation of effective action. This encodes interactions into the fields in a self-consistent manner, and as a result the Landau pole difficulty is resolved. This insight opens a new avenue in study of strong interactions. Therefore, our next step is to study the Savvidy Yang-Mills vacuum state~\cite{Savvidy:1977as, Nielsen:1978rm, Cho:2000ck, Ozaki:2015yja, Savvidy:2022jcr, Savvidy:2023aa}, where the everlasting approach can be applied to the strong field limit of the constant chromomagnetic background.


\end{document}